\documentclass[prl,superscriptaddress,twocolumn,amsmath,amssymb]{revtex4-1}
\usepackage{bm}% bold math
\usepackage{graphicx}%
\usepackage{dcolumn}% Align table columns on decimal point
\usepackage[colorlinks=true]{hyperref}
\usepackage{hyperref}
\hypersetup{
    colorlinks=true,
    linkcolor=blue,
    filecolor=magenta,      
    urlcolor=cyan,
    citecolor=red,
}

\usepackage[version=3]{mhchem} % Formula 
\usepackage{siunitx}
\newcommand{\SP}{Supporting Information}
\newcommand{\F}[1]{Figure \ref{FIG#1}}

\newcommand{\txblRtoBl}[1]{\textcolor{black}{#1}}
\newcommand{\txblBtoBl}[1]{\href{https://pubs.acs.org/doi/10.1021/acs.analchem.0c02884}{\textcolor{blue}{#1}}}

\newcommand{\ED}[1]{\SI{#1}{e^-\,\AA^{-2}\,s^{-1}}}

\begin{document}

% \preprint{Manuscript ID: xxx}

\title%[An \textsf{achemso} demo]
   %{Observation of sub-10-nm thin wall Nanopipette by Transmission Electron microscope}
   {Geometrical Characterization of Glass Nanopipettes with Sub-10 nm Pore Diameter by Transmission Electron Microscopy}

\author{Kazuki Shigyou}
\affiliation%[WPI-NanoLSI]
{Nano Life Science Institute (WPI-NanoLSI), Kanazawa University, Kakuma-machi, Kanazawa 920-1192, Japan}
% \altaffiliation{Contributed equally to this work}
\thanks{Contributed equally to this work}

\author{Linhao Sun}
\affiliation%[WPI-NanoLSI]
{Nano Life Science Institute (WPI-NanoLSI), Kanazawa University, Kakuma-machi, Kanazawa 920-1192, Japan}
% \altaffiliation{Contributed equally to this work}
\thanks{Contributed equally to this work}

\author{Riku Yajima}
\affiliation%[Nano_gakusei]
{Division of Nano Life Science, Graduate School of Frontier Science Initiative, Kanazawa University, Kakuma-machi, Kanazawa 920-1192, Japan}

\author{Shohei Takigaura}
\affiliation%[rikou_butsuri]
{Department of Physics, Institute of Science and Engineering, Kanazawa University, Kakuma-machi, Kanazawa 920-1192, Japan}

\author{Masashi Tajima}
\affiliation%[butsuri]
{College of Science and Engineering, Kanazawa University, Kakuma-machi, Kanazawa 920-1192, Japan}

\author{Hirotoshi Furusho}
\affiliation%[WPI-NanoLSI]
{Nano Life Science Institute (WPI-NanoLSI), Kanazawa University, Kakuma-machi, Kanazawa 920-1192, Japan}

\author{Yousuke Kikuchi}
\affiliation%[rikou]
{Institute of Science and Engineering, Kanazawa university, Kakuma-machi, Kanazawa 920-1192, Japan}

\author{Keisuke Miyazawa}
\affiliation%[front]
{Faculty of Frontier Engineering, Institute of Science and Engineering, Kanazawa University}
%\alsoaffiliation%[WPI-NanoLSI]
\affiliation
{Nano Life Science Institute (WPI-NanoLSI), Kanazawa University, Kakuma-machi, Kanazawa 920-1192, Japan}

\author{Takeshi Fukuma}
\affiliation%[WPI-NanoLSI]
{Nano Life Science Institute (WPI-NanoLSI), Kanazawa University, Kakuma-machi, Kanazawa 920-1192, Japan}
%\alsoaffiliation%[front]
\affiliation
{Faculty of Frontier Engineering, Institute of Science and Engineering, Kanazawa University, Kakuma-machi, Kanazawa 920-1192, Japan}

\author{Azuma Taoka}
\affiliation%[rikou]
{Institute of Science and Engineering, Kanazawa university, Kakuma-machi, Kanazawa 920-1192, Japan}
%\alsoaffiliation%[WPI-NanoLSI]
\affiliation
{Nano Life Science Institute (WPI-NanoLSI), Kanazawa University, Kakuma-machi, Kanazawa 920-1192, Japan}

\author{Toshio Ando}
\email[]{tando@staff.kanazawa-u.ac.jp}
%\phone{+81 (0)76 2645663}
\affiliation%[WPI-NanoLSI]
{Nano Life Science Institute (WPI-NanoLSI), Kanazawa University, Kakuma-machi, Kanazawa 920-1192, Japan}

\author{Shinji Watanabe}
\email[]{wshinji@se.kanazawa-u.ac.jp}
%\phone{+81 (0)76 2344054}
\affiliation%[WPI-NanoLSI]
{Nano Life Science Institute (WPI-NanoLSI), Kanazawa University, Kakuma-machi, Kanazawa 920-1192, Japan}

\begin{abstract}

Glass nanopipettes are widely used for various applications in nanosciences.
In most of the applications, it is important to characterize their geometrical parameters, such as the aperture size and the inner cone angle at the tip region.
For nanopipettes with sub-10 nm aperture and \txblRtoBl{thin} wall thickness, transmission electron microscopy (TEM) must be most instrumental in their precise geometrical measurement.
However, this measurement has remained a challenge because heat generated by electron beam irradiation would largely deform sub-10-nm nanopipettes. Here we provide methods for preparing TEM specimens that do not cause deformation of such tiny nanopipettes.

\end{abstract}

\maketitle

\newcommand*\mycommand[1]{\texttt{\emph{#1}}}
\subsection*{INTRODUCTION}

%\paragraph{}
%Nanopipettes have numerous applications in current nanoscience, such as single molecule monitoring~\cite{yu2017label}, and on-demand delivery~\cite{ivanov2015on-demand}, because of their low cost and easy fabrication.
Glass nanopipettes have numerous applications in nanosciences, such as single molecule sensing of macromolecules and their interactions~\cite{yu2017label} and local delivery of molecules~\cite{ivanov2015on-demand}.
%In the common scheme of single molecule sensing with nanopipettes, the sensitivity and selectivity of detectable molecules, for example DNA and protein molecules, are determined by geometrical factors such as the cone angle and pore diameter of the nanopipette tip~\cite{sze2015fine}.
%!0615差し替え or -> and
%!In many applications (if not all), the nanopipette geometrical parameters, such as the inner and outer cone angles, pore and central channel diameters, or wall thickness are important determinants of the application performance~\cite{sze2015fine}, as typically seen in scanning ion conductance microscopy (SICM)~\cite{hansma1989scanning,novak2009nanoscale,klausen2016mapping,watanabe2017high}.
In many applications (if not all), the nanopipette geometrical parameters, such as the inner and outer cone angles, pore and central channel diameters, and wall thickness are important determinants of the application performance~\cite{sze2015fine}, as typically seen in scanning ion conductance microscopy (SICM)~\cite{hansma1989scanning,novak2009nanoscale,klausen2016mapping,watanabe2017high}.
%In single molecular sensing field, the sensitivity and accuracy of detected molecules, e.g. DNA and protein molecule, is strongly affected by their geometries including tip shape and pore diameter~\cite{sze2015fine}.
% accuracy は何らかの比較できる数字に使用するものだと思う。強いてやるならaccuracy of the selectivityだけど、ここではselectivityだけ足りると思います。affectは悪い方の影響のイメージがあるので変更。
%Pipette geometry is also important for pipette-based scanning probe microscope techniques, in the so-called scanning ion conductance microscope (SICM)~\cite{hansma1989scanning,novak2009nanoscale,klausen2016mapping,watanabe2017high}.
The spatial resolution of SICM is strongly dependent on the nanopipette tip geometry~\cite{rheinlaender2015lateral,rheinlaender2009image,weber2014experimental,edwards2009scanning,rheinlaender2017accurate,del2014contact}.
The assessment of detailed tip geometry is thus important for the optimization of nanopipette-based measurements. Two methods and their combination have been widely used for this assessment: electrical measurements of the ion current resistance and scanning electron microscopy (SEM) imaging~\cite{rheinlaender2017accurate}.
% %In scanning ion conductance microscope (SICM) field, the scanned image resolution has also been examined to be affected by the pore diameter from theories and experiments~\cite{rheinlaender2015lateral}.
% Obtaining the detailed geometry of the tip is thus important for the optimization of nanopipette-based measurements.
% Two methods, and their combination, have been widely used for determining nanopipette tip shape: pipette resistance and scanning electron microscope (SEM) measurements~\cite{rheinlaender2017accurate}.
The ion current resistance is related to the pore diameter and the inner half  cone (IHC) angle~\cite{tognoni2016characterization,perry2016characterization,ying2002prog}.
% The pipette resistance obtained from electrical measurements is related to pipette tip characteristics such as pore diameter and inner half cone (IHC) angle~\cite{tognoni2016characterization}.
    %The electrical conductivity measurement gives the pipette resistance that is related to the pipette tip geometries such as pore diameter and inner half cone angle (IHC) of the pipette~\cite{tognoni2016characterization}.
The IHC angle cannot be directly assessed in both measurements.
    % Usually the pipette IHC angle is estimated from SEM measurements, since electrical measurements do not give this angle directly.
% When the tip shape of fabricated nanopipettes shows good reproducibility, the pipette pore diameter can be estimated from the electrical measurements with the use of the IHC angle estimated from SEM measurements\cite{rheinlaender2017accurate}.
%!0615差し替え the -> a
%!Therefore, it is usually estimated from the outer cone angle given by SEM images of the nanopipette by assuming a constant ratio of outer to inner radius of the nanopipette, or from the pore size given by SEM images and the ion current resistance~\cite{rheinlaender2017accurate}.
Therefore, it is usually estimated from the outer cone angle given by SEM images of the nanopipette by assuming a constant ratio of outer to inner radius, or from the pore size given by SEM images and the ion current resistance~\cite{rheinlaender2017accurate}.
%!0615 the -> a
%!When the nanopipette fabrication can be performed with good reproducibility of a constant geometry, the pore diameter can be estimated from the electrical measurement alone using the IHC angle pre-estimated from SEM measurements for a few nanopipettes.
When nanopipette fabrication can be performed with good reproducibility of a constant geometry, the pore diameter can be estimated from the electrical measurement alone using the IHC angle pre-estimated from SEM measurements for a few nanopipettes.
% Although SEM measurements allow visualization of the pipette outer shape and pore diameter, the metal-coating pretreatment that is usually required for SEM prevents measurement of the internal pore diameter of the pipette.
However, SEM visualizes the pipette outer shape and pore after pretreatment with metal-coating, and therefore, the measured geometrical parameters are affected by the metal-coating. 
% Another disadvantage of SEM is that one cannot visualize internal structures in the tip region of the nanopipette.

Direct observation of the detailed tip geometry of nanopipettes has been achieved using transmission electron microscopy (TEM)~\cite{chen2017characterization,sa2013experiment,perry2016characterization}.
% Direct observation of the tip geometry of nanopipettes has been achieved using the transmission electron microscope (TEM)~\cite{chen2017characterization,sa2013experiment,perry2016characterization}, where the pore diameter and IHC angle can be directly identified and accurately measured.
From their TEM images, the pore diameter, IHC angle, and the internal channel can be directly and accurately measured.
% However, this success in visualization and characterization of nanopipettes is limited to pore diameters greater than \SI{20}{nm}~\cite{perry2016characterization}.
%However, this method application is limited to nanopipettes with a pore diameter of larger than \SI{20}{nm}~\cite{perry2016characterization}.
%! revision from reviewer#2 "limited"
%\txblRtoBl{However, this method is still challenging for pipettes with pore diameter \SI{30}{nm}~\cite{perry2016characterization}, particularly to get consistent images.}
\txblRtoBl{However, it is still challenging to perform TEM measurements for pipettes with pore diameter less than \SI{30}{nm}~\cite{perry2016characterization}, particularly to get their consistent images.}
%nanopipettes with a pore diameter of larger than \SI{20}{nm}~\cite{perry2016characterization}.
%! SI
% For TEM measurements on smaller nanopipettes, deformation of the nanopipette tip, due to heat generated by the electron beam, is a serious problem~\cite{chen2017characterization}.
In TEM measurements of smaller nanopipettes with thin wall, deformation of their tip caused by the electron beam-generating heat is a serious problem~\cite{chen2017characterization}.
% Although we visualized sub-10-nm nanopipettes using TEM in our recent work~\cite{sun2019thermally}, we did not consider the artifacts produced by such deformation.
% Determining the optimal operating conditions for TEM measurements is therefore important, to overcome the heating problem and precisely characterize the geometry of sub-10-nm nanopipettes.
%The optimization of TEM measurements is therefore important to overcome this heating problem and thereby achieve precise characterization of the geometry even for nanopipettes with sub-10-nm diameter (referred to as sub-10-nm nanopipettes hereafter).
%! revision from reviewer#2 "even"
%\txblRtoBl{The optimization of TEM measurements is therefore important to overcome this heating problem and thereby achieve precise characterization of the geometry particularly for nanopipettes with sub-10-nm diameter (referred to as sub-10-nm nanopipettes hereafter).}
\txblRtoBl{The optimization of TEM measurements is therefore important in overcoming this heating problem and thereby achieving precise characterization of the geometry, particularly for nanopipettes with sub-10 nm diameter (referred to as sub-10-nm nanopipettes hereafter).}

In this report, we describe methods to fabricate sub-10 nm nanopipettes with good reproducibility and to characterize their tip geometry with TEM without causing tip deformation.
In addition, we provide an easy method to prepare a large number of tip segments on a TEM grid.
These methods facilitate tuning and optimization of sub-10 nm nanopipette fabrication for practical applications.

    %However, in the practical use of TEM, the irradiation of the electron beam can easily destroy and deform the glass pipette tip due to a thin glass wall around the tip region of the nanopipette~\cite{chen2017characterization}, which obstacles our target for the observation of sub-10-nm nanopipette.
    %An optimal operation condition for TEM measurements is greatly desired to overcome the above problem.
    
    % In this study, we describe how to geometrically characterize the tip shape of sub-10-nm nanopipettes using TEM measurements.
    % We show that heat dissipation is required for TEM visualization of sub-10-nm nanopipettes with no deformation.
    % In addition, we provide an easy method to prepare a large number of nanopipettes for analysis on a TEM grid.
    % This method is useful for tuning and optimizing the tip shapes of sub-10-nm nanopipettes for practical applications.

    %Our method shown here would help to deepen the understand how xxx 付け加えようと思ったが、いらない気もする。
    %The success in geometry characterization of sub-10-nm nanopipette could assist to understand the pore size-/cone angle-affected translocation process occurred in resistance pulse technique as well as contribute to high spatial resolution imaging for nanostructures of biolocical samples in the use of nanopipette-based SICM technique. 
    %この部分には今後の展開に関する内容をもう少し入れる。
    %ここに結論を書く ここで示す方法は有用であることとか。
    
    \begin{figure*}[th]
        \includegraphics{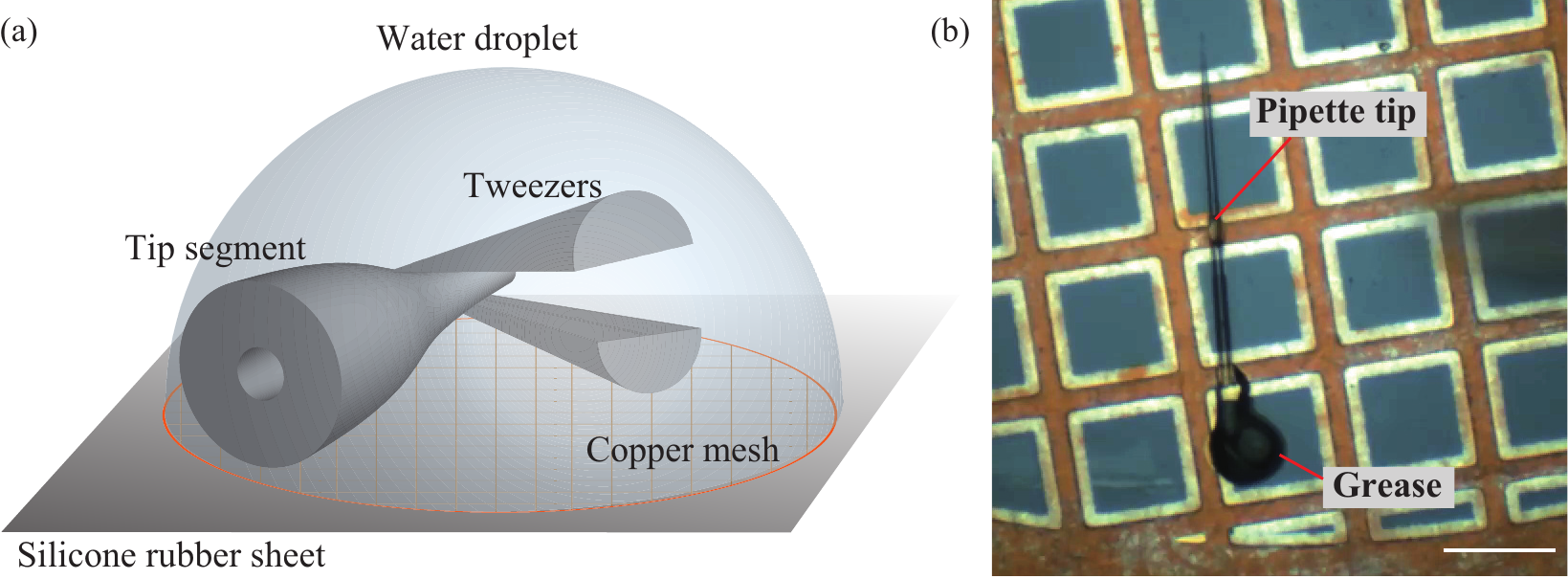}
        \caption{
            (a) Schematic for cutting of a nanopipette to produce a tip segment in a water droplet on TEM grid.
            (b) Optical microscope image showing a tip segment appropriately positioned and glued on the TEM grid. Scale bar, \SI{100}{\micro m}.
        % (a) Pipette is cut in a water droplet on the TEM grid. 
        % (b) Optical microscope image of the pipette fixed with grease as a glue on the TEM grid.
        % Scale bar, \SI{100}{\micro m}
        }
        %ここの書き方は\micro m が基本的な書き方ですか？
        \label{FIG1}
        \end{figure*}
\subsection*{MATERIALS AND METHODS}
    \label{matemetho}

    \subsubsection*{Materials.}
        %!0620 (DI)追加
        \SI{2}{M} KCl and \SI{50}{mM} FeCl$_3$ solutions were prepared using deionized (DI) Milli-Q water (resistivity, $\sim$\SI{18}{M \Omega \cdot cm} at \SI{25}{\degreeCelsius}).
        The FeCl$_3$ solution was used to prepare Ag/AgCl electrodes~\cite{polk2006ag}.
        KCl and FeCl$_3$ were purchased from Funakoshi Co. (Tokyo, Japan) and Sigma-Aldrich (MS, USA), respectively.
        % A \SI{2}{M} KCl solution was prepared by dissolving \SI{7.45}{g} KCl powder (Funakoshi Co., Tokyo, Japan) in \SI{50}{mL} deionized water (DI H$_2$O, resistivity $ \sim$ \SI{18}{M \Omega \cdot cm} at \SI{25}{\degreeCelsius}, Millipore Corp., Danvers, MA, USA).
        % A \SI{50}{mM} FeCl$_3$ solution was prepared by dissolving \SI{1.625}{g} FeCl$_3$ powder in \SI{200}{mL} deionized water (Sigma-Aldrich Co. LLC., MS, USA).
        % This was used to make Ag/AgCl electrodes~\cite{polk2006ag}.
        A flexible syringe needle made of plastic and fused silica (World Precision Instruments, Microfil, MF34G-5) was used to fill nanopipettes with DI water or KCl solution.

    \subsubsection*{Sub-10 nm Nanopipette Fabrication.} 

        All sub-10 nm nanopipettes used in this study were fabricated from quartz capillaries with an inner diameter of \SI{0.3}{mm} and an outer diameter of \SI{1.0}{mm} (Q100-30-1.5, Sutter Instrument, CA, USA).
        The capillaries were cut into \SI{75}{mm} length with a custom-made pipe cutter.
        Nanopipettes were then fabricated with a laser puller (P-2000, Sutter Instrument, CA, USA).
        The details of pulling parameters used for the laser puller are provided in Supporting Information, \txblBtoBl{Table S1}.    
    
    % The nanopipettes were fabricated from a quartz capillary with an inner diameter of \SI{0.3}{mm} and an outer diameter of \SI{1.0}{mm} (Q100-30-1.5, Sutter Instrument, CA, USA).
    %     The capillary was cut into \SI{75}{mm} lengths with a custom-made pipe cutter.
    %     The nanopipettes were then fabricated with a laser puller (P-2000, Sutter Instrument, CA, USA). Details of the pulling parameters of the fabrication program are provided in \textcolor{blue}{Table S1}. %for this paper. 

    \subsubsection*{Current-Voltage Curve Analysis.}
        %!0620 del of the nanopipettes
        %!The current-voltage (I--V) relationship was measured to estimate the nanopipette pore diameter of the nanopipettes before TEM measurements.
        The current-voltage ($I-V$) relationship was measured \txblRtoBl{using the method previously described~\cite{sun2019thermally}} to estimate the nanopipette pore diameter before TEM measurements.
        The nanopipettes were \txblRtoBl{first} filled with DI water by a thermally driven method~\cite{sun2019thermally}.
        %we measured the current-voltage relationship with each pipette.
        %Firstly, the pipettes were completely filled in with the deionized water by a thermal driven method\cite{sun2019thermally}.
        %Then the water was replaced with \SI{2}{M} KCl solution.
        %The I--V measurement was performed in \SI{2}{M} KCl solution.
        %! revision kept for over 10 min
        %\txblRtoBl{Then the water was replaced several times with \SI{2}{M} KCl solution.
        %The I--V measurement was performed in \SI{2}{M} KCl solution.
        %No concentration gradient at the tip due to the replacement was confirmed in the $I-V$ measurement (\SP, SI\,8).}
        \txblRtoBl{Then, a large fraction ($>$98\%) of water inside the pipette was replaced with \SI{2}{M} KCl solution by injecting it.
        The nanopipettes were incubated for \SI{10}{min}.
        For assurance of full solution exchange, the nanopipettes were subsequently immersed in a bath solution of \SI{2}{M} KCl for \SI{10}{min}.
        The absence of a concentration gradient in the tip was confirmed by an $I-V$ measurement over time (\SP, SI\,8).}

        %Secondly, we injected \SI{2}{M} KCl solution to the pipette to replace the water.
        %Finally, ion current flowing through the pipettes was measured in \SI{2}{M} KCl solution.
        Before the measurement, the tip of the pipette was washed with DI water to remove any contamination.
        Ag/AgCl electrodes were prepared by dipping Ag wires into \SI{50}{mM} FeCl$_3$ solution for one min~\cite{polk2006ag}.
        % For the I--V measurements, Ag/AgCl electrodes were prepared by dipping Ag wires into \SI{50}{mM} FeCl$_3$ solution for one minute~\cite{polk2006ag}.
        %Before the measurement, we washed the tip of the pipette using DI water to remove adhered contamination and we dipped Ag/AgCl electrode into \SI{50}{mM} FeCl$_3$ a minute~\cite{polk2006ag}.
        The ion current was amplified with an amplifier (Axopatch 200B, Molecular Devices, LLC, CA, USA).
        % The measured ion current of each pipette was amplified (Axopatch 200B, Molecular Devices, LLC, CA, USA) and the numerical data were transferred to a personal computer (Axon Digidata 1500B, Molecular Devices, LLC., CA, USA).
        A Faraday cage was used to suppress electrical noise in the $I-V$ measurements. A narrow range of bias voltage (between $-40$ to \txblRtoBl{$+$}\SI{40}{mV}) was used, in which ion current rectification was negligible \txblRtoBl{(Figure. S6)}.
        % A Faraday cage was used to effectively suppress electrical noise in the I--V measurements.
        % %以下で良いか確認してください。
        % A narrow range of bias voltage (\SIrange{-40}{40}{mV}) was used to suppress the effects of ion-current rectification.
        % In this range, the I--V curve obtained in the \SI{2}{M} KCl solution  was almost linear. 
        % The electrical resistance of each nanopipette was obtained using MATLAB (MathWorks, Inc., MA, USA) with our custom code and a KCl-solution conductivity of \SI{21.62}{S/m}.
        The pipette resistance (V$/$I) was analyzed as a function of measured pore diameter, IHC angle, and ion conductivity (\SI{21.62}{S/m} at \SI{25}{\degreeCelsius} for \SI{2}{M} KCl) using a lab-made program based on MATLAB (MathWorks, Inc., MA, USA).

        \subsubsection*{Sample Rreparation for TEM Observations.}
        %After I--V measurement, the inner side of those nanopipettes filled with \SI{2}{M} KCl solution was replaced by injection of DI water for three times.
        %!After the I--V measurements, the nanopipettes were emptied of the \SI{2}{M} KCl solution and refilled with DI water three times.
        %
        %!0620 as far as possible ... -> residual ions
        %!After the I--V measurements, the \SI{2}{M} KCl solution in the nanopipettes was exchanged to DI water.
        %These pipettes were then stored overnight in DI water, to remove residual ions.
        After the $I-V$ measurements, the \SI{2}{M} KCl solution in the nanopipettes was exchanged to DI water.
        These pipettes were then stored overnight in DI water to remove residual ions.
        %, as far as possible, any KCl solution remaining in the tip.
        To have a short segment containing the intact tip portion of a nanopipette, the nanopipette was cut in a \SI{12}{\micro L} deionized water droplet on a TEM grid, using a micromanipulator and a micro-tweezers (AxisProSS, Micro Support, Shizuoka, Japan) (Supporting Information, \txblBtoBl{Movie 1}).
        %! revision reviewer2 "making the role of the water droplet clear"
        %\txblRtoBl{The water droplet provides functions of (i) damping to avoid the tip end to be damaged when a cut segment of the pipette is deposited on the TEM grid, and of (ii) traping the cut segment in the water droplet.}
        \txblRtoBl{The water droplet functions as (i) a cushion against damage of the tip end that would otherwise occur when a cut segment is deposited on the TEM grid, and as (ii) a cage that traps the cut segment in the water droplet.}
        % The pipettes were cut to tips in a \SI{12}{\micro L} DI water droplet on a TEM grid.
        % A silicone rubber sheet with a hydrophobic surface (Figure 1a) was used to form the DI water droplet on the TEM grid.
        % After being cut off the pipettes, the tips were effectively trapped in the DI water droplet.
        % The cutting was performed using a micromanipulator with tweezers (\textcolor{blue}{Movie 1}).
        The TEM grid was placed on a hydrophobic silicone rubber sheet (Figure \ref{FIG1}a), which facilitated the formation of a hemisphere-shaped water droplet on the TEM grid.
        %After a number of nanopipettes being cut in this way, the resulting segments were trapped in the water droplet.
        %! revision 
        %\txblRtoBl{After a number of nanopipettes being cut in this way, we confirmed that all the resulting segments were on the TEM grid.}
        \txblRtoBl{After cutting a number of nanopipettes in this way, we confirmed that all the resulting segments were on the TEM grid.}
        After full evaporation of the water droplet, the tip segments were properly positioned on the TEM grid, using the micromanipulator with a probe (Supporting Information, \txblBtoBl{Movie 2}).
        % After the DI water droplet had evaporated, the nanopipettes were aligned and fixed into their proper positions (\textcolor{blue}{Movie 2}).
        %位置移動と固定の部分には、最初にシリコングリースを移動用ピペットに塗布してから切断したピペットを移動したことに関して言及する。移動時に利用するピペットの角度は、30～45度で固定した。
        % TEM copper grids with and without carbon-formvar membranes were used to investigate deformation effects on nanopipette tips in TEM measurements.
        %We prepared the nanopipette samples on two different types of TEM grids for checking the deformation effect.
        TEM copper grids with or without a formvar$/$carbon membrane were used to investigate how the membrane affects the heat-induced tip deformation during TEM measurements.
        
    \subsubsection*{TEM Measurements.}
        The TEM grid on which the nanopipette segments had been properly positioned was placed in the chamber of the TEM apparatus (JEOL 2100 plus).
%
        % The TEM grid on which the nanopipette tips were aligned was placed in the vacuum container of the TEM (JEOL 2100 plus).
%
        TEM measurements were performed at \SI{200}{kV} acceleration voltage.
        % After sufficient time in vacuum, TEM measurements were performed at \SI{200}{kV} acceleration voltage.
        %Before TEM observation, the samples were put into vacuum container for removing evaporable component in the sample.
        %The pipettes were observed by JEOL 2100 plus under \SI{200}{kV} acceleration voltage.
        %位置合わせ時のCDは0.5～0.6pA/cm^2の値で大まかな位置調整を行い、その後MDSを使って露出を1sに固定して観察を行った。ということを追加する。
        %! 200kV -> SI{200}{kV}
        %! 0620 in our -> in the, the tip deformation -> tip deformation
        The minimum dose system (MDS), a measurement mode in the TEM apparatus, was used to minimize tip deformation.
        In the MDS, the electron beam irradiates the sample only when the TEM image is captured.
        The beam irradiation time for capturing an image was set at one second.
        By utilizing the MDS, we were able to precisely control the total amount of electron dose during imaging, which allowed us to examine an effect of heat generated by electron-beam irradiation.
        %! current density -> electron density
        % This enabled us to examine the effects of heat generated by electron-beam irradiation in TEM measurements, as shown in Figures \ref{FIG2} and \ref{FIG3}.
        All TEM images were captured at a magnification range between $\times$40,000 and $\times$80,000.
        % shown in this paper were successively captured at \SI{1}{frame/s}, other than noted, and they have magnifications in the range $\times$40\,000 to $\times$80\,000.
        %!\textcolor{red}{The electron dose of TEM images was estimated from the current density of our apparatus that was proportional to the irradiation electron density.}
        %!0620 簡素化
        The electron dose during image acquisition was estimated from the current density.
        %In this case, a constant current density of \SI{18.2}{pA/cm^2} in Figure \ref{FIG2} and gradually increased current density from 1.6 to \SI{118.5}{pA/cm^2} in Figure \ref{FIG3}.
        %! [pA/cm$^2$] -> \SI{xx}{pA/cm^2}
        %! [pA/cm$^2$] -> \SI{xx}{pA/cm^2}
        
        %古庄さんの計算：確定版　以下の値かける[pA/cm^2]で計測したカレントメータの値でドーズ量が求められる。
        %!４万倍で 43 e-/nmˆ2 (0.4 e-/Aˆ2)，
        %!８万倍で 173 e-/nmˆ2 (1.7 e-/Aˆ2)

        \subsection*{RESULTS AND DISCUSSION}

        \begin{figure*}[th]
            \includegraphics{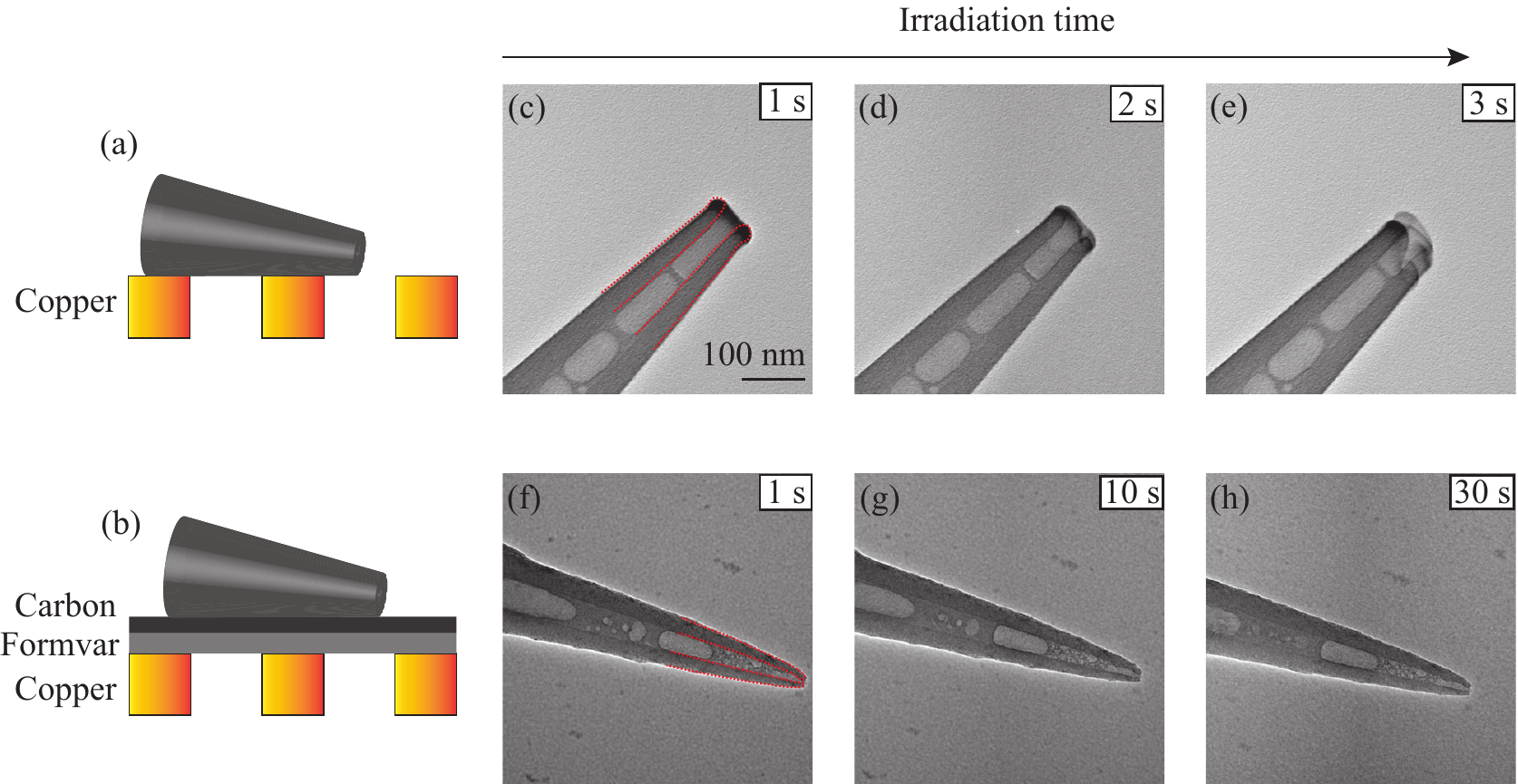}
            \caption{
            (a, b) Schematics of nanopipette tip segments on copper mesh TEM grids without (a) and with the formvar$/$carbon membrane (b).
            (c--h) Time-lapse TEM images of sub-10 nm nanopipettes on the membrane-less TEM grid at 1 to \SI{3}{\second} (c--e), and of those on membrane-coated TEM grid at 1 to \SI{30}{\second} (f--h).
            The broken red lines are a visual guide to the pipette wall outlines. 
            % Schematics of nanopipette tips on TEM grids of copper mesh (a) without and (e) with carbon-formvar membrane.
            % %(a and b) Schematics of nanopipette on TEM grid with and without formvar membrane.
            % %! (a) and (b) -> (a and b)
            % Time-lapsed TEM images of sub-10-nm nanopipettes on copper grid captured for (b--d) \SIrange{1}{3}{\second} without membrane and (f--h) \SIrange{1}{30}{\second} with membrane, respectively.
            % %! the pipette on a copper grid -> pipettes on the copper grid
            % %! within 3 sec respectively -> from 1 to 3 sec.
            % %! (c) to (e)
            % %(f-h) Time-lapse TEM images using TEM grid with formvar membrane from 1 to \SI{30}{sec}
            % %! (f) to (h) -> (f-h)
            % Broken red lines are a visual guide to indicate the pipette walls around the tip.
            %Scale bar in (c), \SI{100}{nm}.
            Note that nano-bubble-like structures are seen in the TEM images (see \SP, Section \txblBtoBl{SI\,5} and \txblBtoBl{Movie 3}).
            }
            \label{FIG2}
            \end{figure*}  

        %Subsectionに分けたほうがよい 200217
        \subsubsection*{Fabrication of Nanopipettes.}
        Sub-10 nm nanopipettes were fabricated using the laser puller P-2000, in which a \SI{75}{mm} long glass capillary was clamped at their ends.
        The parameters for the puller action listed in Supporting Information, \txblBtoBl{Table S1}, were used. We noticed that the precise control of heat dissipation during the pulling process was important for producing sub-10 nm pipettes with high reproducibility.
        The rate and amount of heat dissipation are sensitive not only to the area and position at which the glass capillary is clamped but also to the clamping force.
        The silicone rubber used to clamp a glass capillary deforms depending on the magnitude of clamping force.
        Therefore, the shape of a produced nanopipette is affected by the clamping force.
        To control the clamping force precisely, we used a torque driver and a custom screw head.
        Details are provided in Supporting Information, Section \txblBtoBl{SI\,2}.
%
        % Sub-10-nm nanopipettes were fabricated using laser puller P-2000 with the pulling program listed in \textcolor{blue}{Table S1}.
        % We found that control of the exhaust heat for the pulling process was important for producing sub-10-nm pipettes with high reproducibility.
        % This exhaust heat is sensitive not only to the contact area clamping the glass capillary, and its position, but also the puller’s clamping force.
        % The silicone rubber of the clamps, used as support and fixation for the glass capillary, deforms by an amount depending on the size of clamping force. Pulling results are therefore affected by the clamping force. 
        % To control and improve the stability of the magnitude of this applied force, we precisely controlled the torque of the screw that determined the force, through the combination of torque driver and custom screw head.
        % In our pulling setup the \SI{75}{mm} glass capillaries  were clamped at their ends.
        % Details are provided in Supporting Information, section \textcolor{blue}{SI\,2}. 
        % 
        %Here, one important improvement in getting relatively stable geometry was performed as shown in Figure S1.
        %We modified the original screw using hexagon head screw, where the constant force can be applied onto the glass capillary.
        % cascrew?とはなんですか？
        %This improvement can solve the difficulty in the control of capillary tightness.
    
    \subsubsection*{Alignment of Nanopipette Segments on TEM Grid.}
    Figure \ref{FIG1} shows how a short tip segment was prepared on a copper mesh TEM grid. Using a micro tweezers, a nanopipette whose back side was held was cut in a water droplet on the TEM grid supported on a silicone rubber sheet (Figure \ref{FIG1}a).
    The water droplet prevented the resulting tip segment from escaping from the TEM grid and also from damage, allowing us to efficiently prepare a large number of segments on a TEM grid (Supporting Information, \txblBtoBl{Movie 1}).
    After the water droplet evaporated, the segments remained on the grid but their position and orientation unavoidably changed; their tip portions were often superposed on the copper mesh.
    We used the micromanipulator with a glass probe to move each segment so that its tip end portion was not superposed on the copper mesh.
    Before the positional adjustment of a segment, it was necessary to put a small amount of silicone grease on the back end of the segment.
    The segment would otherwise be knocked off the TEM grid because of electrostatic interactions between the segment and the glass probe.
    Figure \ref{FIG1}b shows a magnified image of a tip segment on a TEM grid after its positional adjustment.
    The entire process of this positional adjustment is provided in Supporting Information, \txblBtoBl{Movie 2} and Section \txblBtoBl{SI\,3}.    

    \subsubsection*{Importance of Membrane on TEM Grid.}
        Here we describe the importance of the use of a formvar$/$carbon-coated grid in TEM measurements of the sub-10 nm nanopipette tip.
        %!0620 is -> must be more susceptible
        %!The tip portion of a sub-10-nm nanopipette segment is more easily deformed by electron beam-generating heat, compared to a tip with a larger pore and a thicker wall.
        The tip portion of a sub-10 nm nanopipette segment must be more susceptible to electron beam-generating heat, compared to a tip with a larger pore and a thicker wall.
        % Here we describe the importance of the selection of the TEM grid (with \textit{versus} without a membrane) in TEM measurements for sub-10-nm nanopipettes.
        % In general, the tip regions of sub-10-nm nanopipettes are more easily deformed by heat generated by the electron beam than those with larger pores. 
        % It is well known that using a membrane on the TEM grid suppresses heat damage of samples in TEM measurements.
        % We investigated how sub-10-nm nanopipettes deformed for TEM measurements with and without a membrane made from carbon and formvar.
        %!0620
        %!\textcolor{red}{In TEM measurements, the use of a membrane-coated TEM grid is known to suppresses heat damage of specimens qualitatively. However, the effectiveness of the suppression, indeed, depends on specimens used.}
        In TEM measurements, the use of a membrane-coated TEM grid is known quantitatively to suppress heat damage of specimens.
        %!0620
        %!However, the effectiveness of the suppression, indeed, depends on specimens used.
        However, the effectiveness of the suppression largely depends on the specimen used.
        %!0620 deformed -> would deform
        We investigated how the tip segment of a sub-10-nm nanopipette would deform during its TEM measurements using TEM grids with and without a formvar$/$carbon membrane.
            \begin{figure}[th!]
                \includegraphics[width=6.6cm]{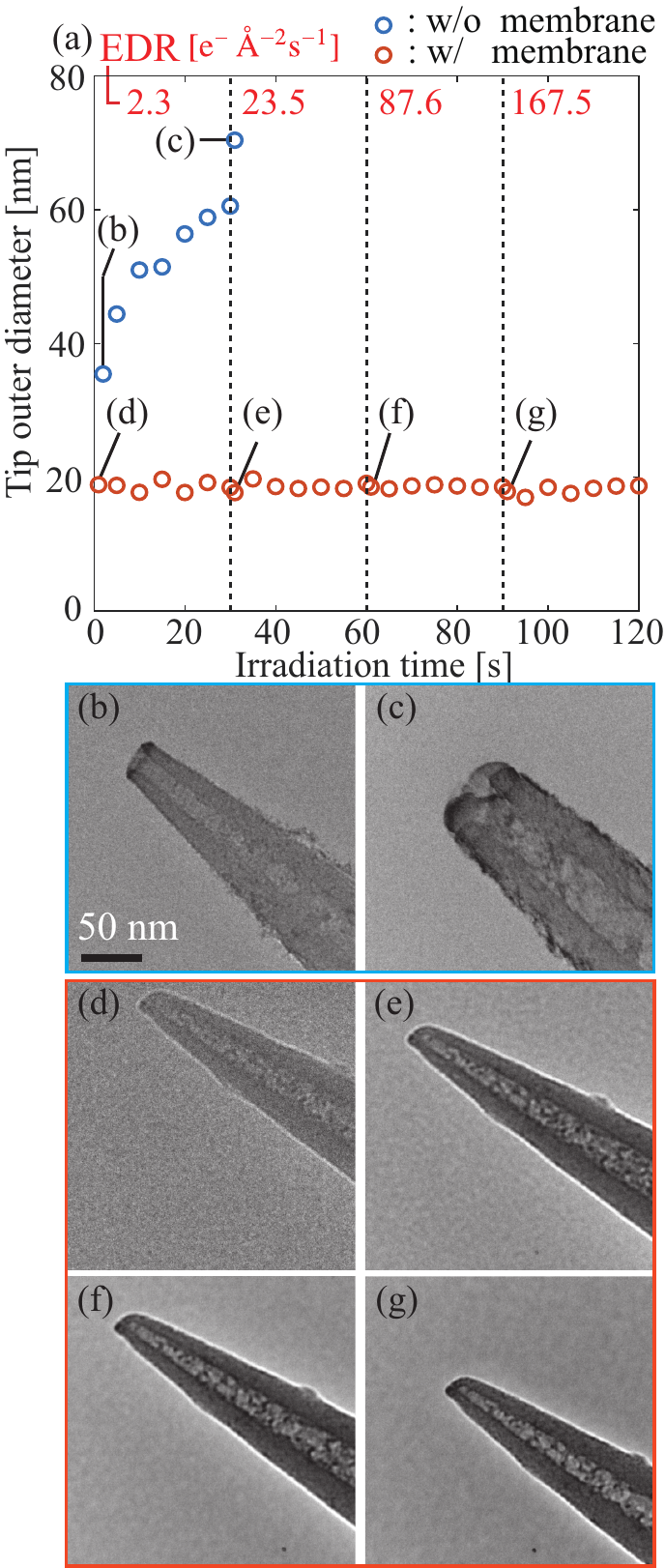}
                \caption{
                Protective effect of membrane-coating of the TEM grid against deformation of the nanopipette tip in TEM imaging.
                (a) Tip outer diameter at the tip end as a function of electron irradiation time and electron dose rate, EDR. The red and blue circles represent outer tip diameters measured from time-lapsed TEM images captured using TEM grids with and without the membrane, respectively.
                The EDR was increased in a stepwise manner from \SIrange[range-units = single]{2.3}{167.5}{e^-\,\AA^{-2}\,s^{-1}} during the TEM measurements.
                (b--g) TEM images captured at the time points indicated by letters corresponding to those shown in (a), using TEM grids without the membrane (b,c) and with the membrane (d--g).
                    %Scale bar in (b), \SI{50}{nm}.
                % (a) Impact of the TEM grid membrane on deformation of nanopipette tips in TEM measurements.
                % Red and blue circles represent outer diameters of nanopipette tips measured from time-lapsed TEM images obtained for TEM grids with and without the membrane, respectively.
                % %The current density increased in a stepwise manner from \SI{1.6}{pA/cm^2} to \SI{118.5}{pA/cm^2} during the TEM measurements.
                % The electron dose (ED) increased in a stepwise manner from \textcolor{blue}{\SI{4.2e2}{e^-/nm^2\,s} to \SI{3.1e4}{e^-/nm^2\,s}} during the TEM measurements.
                % %The nanopipette tip on TEM grid without the membrane was broken at $t$ = \SI{32}{sec}, indicated as the red cross.
                % Each TEM image (b)--(g) was captured at the time indicated by the corresponding letter on the plot.
                % Scale bar in (b): \SI{50}{nm}.
                }
                \label{FIG3}
                \end{figure}

        %how to optimize the condition for TEM observation in order to avoid the deformation of nanopipette tip caused by high density of transmission electrons.
        % Figure \ref{FIG2} shows time-lapsed TEM images of nanopipettes using TEM grids with and without the membrane.
        
%!0703 del:from tow nanopipettes, 
        Figure \ref{FIG2} shows time-lapse TEM images of tip segments placed on TEM grids with and without the membrane.
        %!0703 add
        These segments were obtained from \txblRtoBl{two} nanopipettes fabricated under an identical condition.
        %! 0620 taken using -> placed on 
        % Figure \ref{FIG2} shows time-lapse TEM images of tip segments from two nanopipettes placed on TEM grids with and without the membrane.
        %!An irradiation current density of \SI{18.2}{pA/cm^2} was used to produce these images.
        These TEM images were captured at \SI{1}{frame/s} with an electron dose rate (EDR) of \ED{25.7}.
        %\textcolor{red}{\SI{4.7e2}{e^-/nm^2\,s}}.
        %あとでちゃんとlinkをつける 200217
        % For TEM images without the membrane (Figure \ref{FIG2}a), deformation of the tip region was observed within a few seconds (Figure \ref{FIG2}b--d).
        %The TEM images taken without the membrane (Figure \ref{FIG2}a) show noticeable deformation of the tip region at $t$ = 2 and \SI{3}{\second}, compared to the image at $t$ = \SI{1}{\second}.
        %! revision from reviewer2 "clear damage was observed even at t = 1s"
        %\txblRtoBl{The TEM images taken without the membrane (Figure \ref{FIG2}a) show noticeable deformation of the tip region even at $t$ = \SI{1}{\second}. This deformation becomes significant for $t$ = 2 and \SI{3}{\second}.}
        \txblRtoBl{The TEM images taken without the membrane (Figure \ref{FIG2}c--e) show noticeable deformation at the tip region even at $t$ = \SI{1}{\second}. This deformation becomes significant at $t$ = 2 and \SI{3}{\second}.}
        % In contrast, for TEM images with the membrane (Figure \ref{FIG2}e), no clear deformation was found even at $t$ = \SI{30}{s} (Figure \ref{FIG2}f--h).
        In contrast, the TEM images taken with the membrane (Figure \ref{FIG2}f--h) show no discernible deformation even at $t$ = \SI{30}{s}.
        %!
        %!0620 confidential -> confidence 0703 this -> the
        Note that the pore diameter of nanopipettes subjected to the TEM observations are expected to be within \SI{7.1}{nm} $\pm$ \SI{1.1}{nm} (SD, $n = 8$, as described later) before the TEM measurements.
        %which can be calculated from Figure \ref{FIG4}b).
        %!\textcolor{red}{Note that the two nanopipettes subjected to this TEM observations were twins produced from single pulling, so that their original tip shapes must have been nearly identical before the TEM measurements.}
        %!正しいのか確認すべし
        %!正しくない；同じパラメータつかったくらいしかない。ここで確からしさは、図４からバラツキでみるのがよいと思います。\SI{7.1}{nm}$\pm$\SI{1.1}{nm}(SD, $n = 8$)そうなっている。
        %
%! 自由度n=8 -(1),不偏分散から, alpha=95, 2.3646 (n=7), s^2 = 8.2206,%! 内直径の95％信頼区間(nm)：xbar(7.1129)$\pm$2.397
%! 形の確かさは 7.1$\pm$2.4        %
        %
        %!0620 this -> this original identical size
        We had confirmed this original identical size by measuring ion conductance of these two nanopipettes before their TEM observations.
        Therefore, the tip shape had already been deformed into a trumpet-like shape at $t$ = \SI{1}{\second}, when the membrane-less TEM grid was used.
        In fact, when a membrane-less TEM grid was used, we always observed the trumpet-like shape, even under electron irradiation with a lower EDR of \ED{2.3} (\SP, Section \txblBtoBl{SI\,4}, \txblBtoBl{Figure S3}).
        In contrast, when a membrane-coated TEM grid was used, a semi-conical tip shape was always observed. 
        We further evaluated the heat-induced deformation and the membrane effect against this deformation by varying the irradiation electron density in the TEM measurements (Figure \ref{FIG3}).
        % To gain more insight into the deformation, we evaluated the effect of varying the irradiation electron density in the TEM measurements (Figure \ref{FIG3}).
        %where two different types of grids were used.
        %In this experiment, the outer diameters of the nanopipette tips were measured as the current density (CD), which is proportional to the irradiation electron density, was changed in a stepwise manner from \SI {1.6}{pA/cm^2} to \SI {118.5}{pA/cm^2}.
        %! 古庄さんのコメントで修正 CDは一般的でない。 -> the irradiation electron density. the electron dose in an image (\SI{xxx}{e^-/nm^2}) 
        In this evaluation, the outer diameter of the tip end was measured as a function of ED.
        The EDR was changed in a stepwise manner from 2.3 to \ED{167.5}.
%
        % In this experiment, the outer diameters of the nanopipette tips were measured as the electron dose (ED) was changed in a stepwise manner from \textcolor{red}{2.3 to \SI{167.5}{e^-/\AA^2\,s}}.
        %以下は修正の必要がある 200214
        %3つというのは、どこできくのでしょうか？よく理解できないのであとでききます。
        %解決
        When a membrane-less TEM grid was used, the tip outer diameter increased even under the condition of EDR = \ED{2.3}, from \SI{35.5}{nm} (Figure \ref{FIG3}b) to \SI{44.4}{nm} during the period from $t$ = \SIrange{2}{5}{s} (the first two blue circles in Figure \ref{FIG3}b).
        This large increase was followed by a gradual increase up to \SI{60}{nm} at $t$ = \SI{30}{s} (the next five blue circles in Figure \ref{FIG3}a).
        When the EDR was increased from 2.3 to \ED{23.5}, the tip outer diameter suddenly increased from \SIrange[range-units = single]{60}{70}{nm} at $t$ = \SI{31}{s} (Figure \ref{FIG3}c).
        In contrast, when a membrane-coated TEM grid was used, there was no discernible deformation of the tip even at EDR = \ED{167.5}, the maximum EDR value for our TEM system (Figure \ref{FIG3}d--g).
        % For nanopipettes without the membrane, at EDR = \SI{2.3}{e^-/\AA^2\,s} the tip outer diameter gradually increased from \SIrange{42}{60}{nm} at $t$ = \SI{30}{s}.
        % As EDR was increased from \textcolor{blue}{2.3 to \SI{167.5}{e^-/\AA^2\,s}}, the tip outer diameter suddenly increased at $t$ = \SI{31}{s} (TEM image in Figure \ref{FIG3}c).
        % In contrast, for nanopipettes with the membrane, there was no clear deformation of the tip even at EDR = \textcolor{blue}{\SI{167.5}{e^-/\AA^2\,s}}, the maximum EDR value for our TEM system.
        %!The tip outer diameter remained $\sim$\SI{17}{nm} throughout the observations (red circles in Figure \ref{FIG3}a), indicating effective dissipation of electron beam-generating heat through the membrane.
        %!The tip outer diameter remained $\sim$ \txblRtoBl{\SI{18.5}{nm}} throughout the observations (red circles in Figure \ref{FIG3}a\txblRtoBl{; \SI{18.65}{nm} $\pm$ \SI{0.73}{nm}(SD), \SI{18.62}{nm} $\pm$ \SI{0.65}{nm}(SD), \SI{18.61}{nm} $\pm$ \SI{0.19}(SD){nm}, and \SI{18.10}{nm} $\pm$ \SI{0.65}{nm}(SD); $n = 7$ for \ED{2.3}, \ED{23.5}, \ED{87.6}, and \ED{167.5}}), indicating effective dissipation of electron beam-generating heat through the membrane.
        % n = 7を消した。
        The tip outer diameter remained $\sim$\SI{18.5}{nm} throughout the observations (red circles in Figure \ref{FIG3}a; 18.65 $\pm$ \SI{0.73}{nm} (SD), \SI{18.62}{nm} $\pm$ \SI{0.65}{nm} (SD), 18.61 $\pm$ \SI{0.19} (SD){nm}, and \SI{18.10}{nm} $\pm$ \SI{0.65}{nm} (SD) for 2.3, 23.5, 87.6, and \ED{167.5}, from left to right), indicating effective dissipation of electron beam-generating heat through the membrane.
        %On formvar-based membrane grid, the out diameter of nanopipette end remains a constant value of \SI {17}{nm} no matter the change of electron intensity from \SI{1.6}{pA/cm^2} to \SI{118.5}{pA/cm^2}, which suggests no deformation occur on such a grid.
        %However, for nanopipette tip supported by free formvar-based membrane.
        %The pipette tip was gradually deformed and broken even under the weakest electron intensity of \SI{1.6}{pA/cm^2}, indicated from the increased out diameter of nanopipette tip end. 
        %! delete: completely
        % We confirmed that 26 nanopipette tips showed no deformation when using the TEM grid with membrane (not shown).
        We confirmed this absence of deformation for 26 sub-10 nm nanopipette tips (not shown).
        In Supporting Information, \txblBtoBl{Table S2}, we provided statistics for tip geometry of eight nanopipettes fabricated using the same pulling parameters, as shown in \SP, \txblBtoBl{Table S1}.
        %Duringはその期間のどこかの点（複数でもよい）、Forはその期間ずっと、ということを意味します。
        %!To confirm it, the results in fig. S1 give measured TEM images for 26 sub-10-nm nanopipettes.
        The tip outer diameter assessed using membrane-coated TEM grids ranges between 14.43 and \SI{26.05}{nm} (mean value, \SI{19.06}{nm}), as shown in \SP, \txblBtoBl{Table S2}.
        %!安藤先生の文章だけど、値の根拠がよくわからない。
        These values are much smaller than the value, \SI{35.5}{nm}, observed at $t$ = \SI{2}{s} using a membrane-less TEM grid (\F{3}b), suggesting that the heat-induced deformation is significant even with 1--2 s exposure to electron beam (dilation rate, $\sim$\SI{5}{nm/s} at EDR = \ED{2.3}).
        %\SI{2.3}{e^-/\AA^2\,s}).
        %Figure \ref{FIG3}b
        Thus, the use of membrane-coated TEM grids is essential for the accurate TEM assessment of tip geometry for sub-10 nm nanopipettes.

    \subsubsection*{Geometrical Characterization of sub-10 nm Nanopipettes: TEM $Versus$ $I-V$ Measurements.}

    \begin{figure*}[ht!]
        \includegraphics{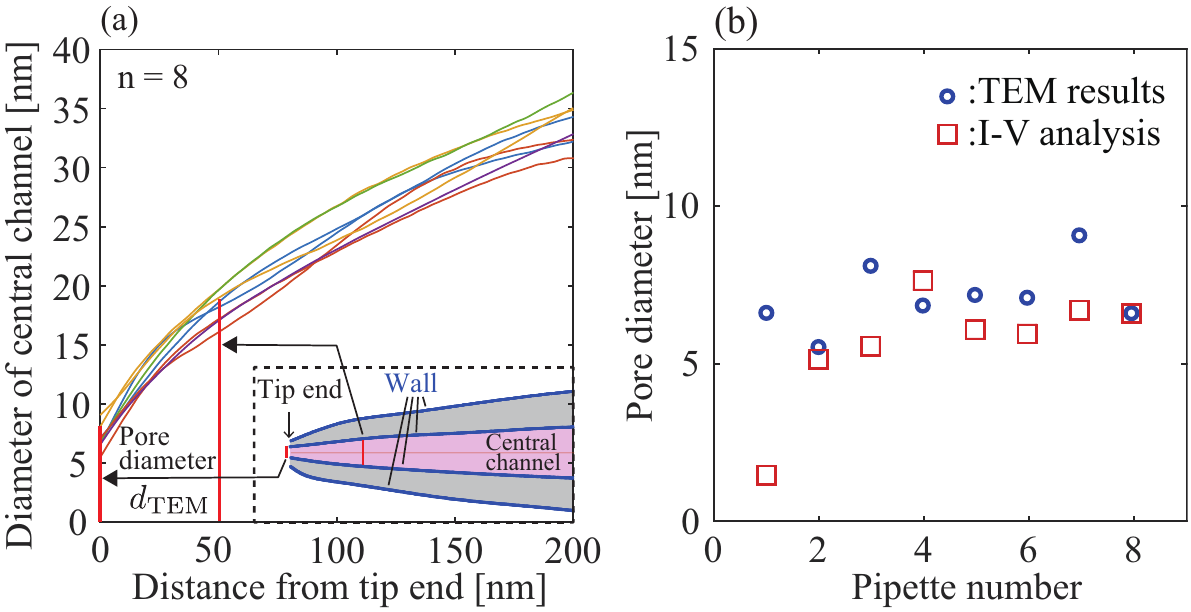}
        %\caption{\label{FIG4} Fitting result by Del Linz equation based on pore radius and half cone angle from TEM results as well as pipette resistance from {\it{I--V}} curve measurement.}
        \caption{
        %TEM results of pipettes shape (a), the schematic image which is inside black dot line shows examples of diameter each distance from tip.
        % (a) Inner diameter of nanopipette tip measured as a function of distance from the nanopipette-tip end, captured from TEM data.
        % The inset indicates schematically how the diameter was extracted from TEM data.
        % Vertical red lines show the corresponding diameter in the TEM data and the schematic.
        % (b) Comparison of pore diameters measured directly from TEM data (blue circles) and estimated from I--V measurements (red circles).
        (a) Diameter of the tip central channel of eight nanopipettes measured from their TEM images as a function of distance from the tip end. The inset indicates schematically how the diameter was extracted from a TEM image.
        The vertical red lines show the diameters corresponding to those shown in the schematic.
        (b) Comparison of pore diameters measured directly from TEM images of eight nanopipettes (blue circles) to those estimated from $I-V$ measurements for the same nanopipettes (red \txblRtoBl{square}). 
        %Scale bar, \SI{100}{nm}.
        }
        \label{FIG4}
        \end{figure*}

        Finally, we discuss how the tip geometry of sub-10-nm nanopipettes is related to the pipette resistance $R_p$. 
        The $R_p$ value obtained by electrical measurements is related to the pore diameter $d_{pore}$ and IHC angle $\theta$ of the nanopipette, and the solution conductivity $\kappa$ through the equation
        \begin{equation} \label{Del_Linz}
        d_{pore}=\frac{2}{\kappa \,R_{p} }(\frac{1}{\pi \tan \theta}+\frac{1}{4}).
        \end{equation}
        % This simple model~\cite{del2014contact} assumes that the pipette has a simple conical shape.
        This model~\cite{del2014contact} assumes a simple conical shape for the tip (i.e., a constant $\theta$).
%!
%!0620 eight -> eight tip segments
        However, the tip regions of sub-10-nm nanopipettes directly captured by TEM revealed deviations from a simple conical shape. 
        Figure \ref{FIG4}a depicts the diameter of the inner channel as a function of the distance measured from the tip end (tip distance), extracted from TEM images of eight tip segments with $d_{pore} <$ \SI{10}{nm} (see Supporting Information, sections \txblBtoBl{SI\,6} and \txblBtoBl{SI\,7}).
        In all cases the diameter does not increase linearly with the tip distance, that is, the IHC angle $\theta$ does not have a single value around the tip, as reported for nanopipettes with larger pore diameters~\cite{tognoni2016characterization}.
        To check how largely the TEM-assessed geometry of a nanopipette tip affects Equation (\ref{Del_Linz}), we used a linearly fitted inner half cone (LF-IHC) angle, defined as the slope obtained by linear fitting of each curve, as shown in \F{4}a for the tip distance from 0 to \SI{200}{nm}(i.e., approximated IHC angle).
        For the eight nanopipettes, we compared the values of $d_{pore}$ measured from their TEM images 
        ($d_{\textrm{TEM}}$) with those estimated using Equation (\ref{Del_Linz}), where $\theta$ was approximated by the LF-IHC angles (\SP, \txblBtoBl{Table S2}).
        The results are shown in \F{4}b.
        % Except for the case of nanopipette number 1, the two sets of values distributed  around \txblRtoBl{$\sim$\SI{7.0}{nm}}, and their respective difference were within \txblRtoBl{\SI{2.9}{nm} ($n = 7$)}, which is 95\% confidence interval.
        %!; reference~~$\sim$\SI{6.7}{nm} $\pm$\SI{2.1}{nm}(SD, $n = 8$)}.
        %!n=7(電気抵抗基準) \SI{0.3}{nm}$\pm$\SI{1.4}{nm}
        %!n=8 \SI{0.3}{nm}$\pm$\SI{1.4}{nm}
        %![この数値をきちんと計算して].
        %!以下はn=8,こっちの方が自然。
        %Except for the case of nanopipette number 1, the two sets of values, $d_{pore}$ and $d_{\textrm{TEM}}$, distributed around $\sim$\SI{7.0}{nm}, and their respective difference were within $\pm$\SI{1.4}{nm} (SD, $n = 7$); therefore, $d_{pore}$ \SI{-1.7}{nm} $\leq \overline{d_{\textrm{TEM}}} \leq d_{pore} +$ \SI{1.1}{nm} (95\% confidence interval, $n=7$), where $\overline{d_{\textrm{TEM}}}$ is the population mean of $d_{\textrm{TEM}}$.
%! revision n=7->n=8の方が自然
%\txblRtoBl{The two sets of values, $d_{pore}$ and $d_{\textrm{TEM}}$, distributed around $\sim$\SI{6.7}{nm}, and their respective difference were within $\pm$\SI{2.1}{nm} (SD, $n = 8$); therefore, $d_{pore}$ \SI{-0.8}{nm} $\leq \overline{d_{\textrm{TEM}}} \leq d_{pore} +$ \SI{2.6}{nm} (95\% confidence interval, $n=8$), where $\overline{d_{\textrm{TEM}}}$ is the population mean of $d_{\textrm{TEM}}$.}
\txblRtoBl{The two sets of measured values of $d_{pore}$ and $d_{\textrm{TEM}}$, distributed around $\sim$\SI{6.4}{nm}, and their difference were within $\pm$\SI{1.5}{nm} (SD, $n = 8$); therefore, a relation of \txblRtoBl{$\overline{d_{pore}}$ \SI{-0.1}{nm} $\leq \overline{d_{\textrm{TEM}}} \leq \overline{d_{pore}} +$ \SI{3.1}{nm} holds (95\% confidence interval, for $n=8$), where $\overline{d_{\textrm{TEM}}}$ and $\overline{d_{pore}}$ are the population mean of $d_{\textrm{TEM}}$ and $d_{pore}$, respectively.}}

        %\txblRtoBl{$\sim$\SI{6.7}{nm}}, and their respective difference were within \txblRtoBl{$\pm$\SI{2.1}{nm}(SD, $n = 8$)}.
        %!
        % For the 8 nanopipettes we compared tip diameters measured from TEM data with those estimated using Equation (\ref{Del_Linz}), in which $\theta$ was the LF-IHC angle (\textcolor{blue}{Table S2}).
        % This is shown in Figure \ref{FIG4}b.
        % Although the tip diameters estimated using Equation (\ref{Del_Linz}) were in the range \SIrange{1.6}{8.6}{nm}, the differences between the diameters obtained from TEM data and Equation (\ref{Del_Linz}) were less than one order of magnitude.
        % This indicated the validity of using electrical measurements to estimate the pore diameter of sub-10-nm nanopipettes.
        
        %IHCの変化がどういうことに関係するのか議論するべき。
        % Although the LF-IHC angle gives a good estimation of the pore diameter, knowing the variation in IHC angle near the tip is still important for the numerical analysis of nanofluidic behavior in nanopipettes, since this analysis requires details of the tip geometry.
        Although the LF-IHC angle gives a good estimate of $d_{\textrm{TEM}}$, knowing the variation in the IHC angle in the tip region is still important for the numerical analysis of nanofluidic behavior in nanopipettes, since this analysis requires details of the tip geometry.
        Finite element analysis simulations on the basis of the coupled Poisson, Nernst--Planck, and Naiver--Stokes equations have been widely used for the analysis of nanopipettes.
        This yields spatial profiles of physical quantities (such as potential, ion concentration, flow velocity and pressure) around the nanopipette tip. Such spatial information is difficult to obtain experimentally.
        Nanofluidic behaviors can be analyzed experimentally for nonlinear $I-V$ relationships, which are generally found in nanofluidic systems with broken symmetry, \txblRtoBl{such as nanopores~\cite{siwy2006ion,siwy2004nanodevice,cheng2007rectified,yusko2010electroosmotic,cao2011concentration,qiu2018abnormal,siwy2002fabrication,woermann2003electrochemical,cervera2005poisson,white2008ion,lan2016voltage} and nanopipettes~\cite{wei1997current,sa2011rectification,deng2014effect}}.
        \txblRtoBl{The combined analysis makes it possible not only to improve the quantitative interpretation of experimental data obtained with nanopipettes but also to extend application of nanopipettes.  The TEM method will make significant contributions to improving, for example, nano-scale surface charge density (SCD) mapping with SICM~\cite{klausen2016mapping,fuhs2018direct,perry2016surface,perry2015simultaneous,page2016fast,mckelvey2014surface}.
        The importance of accurate characterization of the tip geometry for SCD measurement was pointed out in previous TEM study~\cite{perry2016characterization}.
        In this application, knowing accurate tip geometries is necessary to quantify the SCD map of the sample because the quantification is sensitive to tip geometries.
        In fact, quantitative SCD measurements with sub-10 nm spatial resolution is still a challenge~\cite{klausen2016mapping} because of the difficulty of characterizing the geometrical parameters of sub-10 nm nanopipettes. 
        Besides, our TEM method can provide an opportunity to perform SICM measurements in complicated situations.
        A recent study~\cite{rabinowitz2019nanoscale} revealed a significant contribution of electroosmotic flow and the SCD of the tip to the measured ion current when a large ion concentration gradient exists around a small aperture of the tip.
        To quantitatively analyze these contributions to the measured SICM data obtained under such conditions~\cite{watanabe2019development}, information of the detailed tip geometry is indispensable.}   

        %\txblRtoBl{However, the inadequate quantitative understanding of complicated transport phenomena could produce signal artifacts in SICM measurements}
        %In SICM measurements, such conditions can improve the spatial resolution and sensitivity~\cite{watanabe2019development}, but inadequate quantitative understanding of nanofluid behavior could produce signal artifacts near the surface of soft objects.
        
        %!The validity of estimating the nanopipette aperture from the LF-IHC angle breaks down when the nanopipette tip deviates further from a conical shape, like an hourglass-shape nanopipette~\cite{perry2016characterization,Zweifel2016helium,chen2017characterization,rabinowitz2019nanoscale,Holub2020single,Niya2013experiment,Hao2016nanopipette,Yuill2013electrospray,Zhu2020low}.
        The trumpet-like shaped nanopipettes have been frequently observed in electron microscopy measurements~\cite{perry2016characterization,Zweifel2016helium,chen2017characterization,rabinowitz2019nanoscale,Holub2020single,Niya2013experiment,Hao2016nanopipette,Yuill2013electrospray,Zhu2020low}. 
        However, in this study, we were unable to find appropriate pulling conditions that could produce trumpet-liked shaped nanopipettes.
        This suggests that the trumpet-like shaped nanopipettes are resultant from tip deformation by the electron beam (see \SP, \txblBtoBl{Figures S3} and \txblBtoBl{S5}), at least, in TEM measurements of sub-10 nm nanopipettes.
        
        % The validity of estimating the nanopipette aperture from the LF-IHC angle breaks down when the nanopipette tip deviates further from a conical shape, like an hourglass-shape nanopipette~\cite{perry2016characterization,Zweifel2016helium,chen2017characterization,rabinowitz2019nanoscale,Holub2020single,Niya2013experiment,Hao2016nanopipette,Yuill2013electrospray,Zhu2020low}.
        % However, we were not able to find appropriate pulling conditions to fabricate such an hourglass-shape nanopipette\txblRtoBl{, which could be resultant from tip deformation by electron beam} (\SP, \txblRtoBl{Figure S3}).

        %! as -> when;　安藤先生が変更。ただ、as か if だと思う。
        %!We did not check this breakdown, since we could not find appropriate pulling conditions to fabricate such an hourglass-shape nanopipette, without deformation by the irradiating electron beam (\textcolor{blue}{Figure S5}). 
    
\subsection*{CONCLUSIONS}

    We have provided the conditions for fabricating sub-10 nm nanopipettes with good reproducibility and procedures for their geometrical characterization by TEM.
    The TEM imaging following our methods enables quick assessment of tip geometry for many nanopipettes, without deformation of the tip by electron beam-generating heat.
    Our methods will assist and facilitate the production of smaller nanopipettes and thereby contribute to improving various nanopipette applications.

\subsection*{ASSOCIATED CONTENT}
    \subsubsection*{Supporting Information}
    The Supporting Information is available free of charge at \url{https://pubs.acs.org/doi/10.1021/acs.analchem.0c02884}.
    Pulling parameters (\txblBtoBl{SI\,1} and \txblBtoBl{Table S1}), optimized pulling conditions (\txblBtoBl{SI\,2} and \txblBtoBl{Figure S1}), photograph of tip segments of sub-10-nm nanopipettes on TEM grid (\txblBtoBl{SI\,3}, \txblBtoBl{Figure S2}), trumpet-like shaped nanopipettes observed under different conditions (\txblBtoBl{SI\,4}, and \txblBtoBl{Figure S3}), dynamics of nanobubbles in sub-10 nm nanopipettes (\txblBtoBl{SI\,5} and \txblBtoBl{Figure S4}), TEM images of sub-10 nm nanopipettes taken using the membrane-coated TEM grid (\txblBtoBl{SI\,6} and \txblBtoBl{Figure S5}), geometrical parameters obtained from the TEM images (\txblBtoBl{Table S2}), \txblRtoBl{$I-V$ traces of the nanopipettes before characterization by TEM imaging (\txblBtoBl{Figure S6})}, and procedures for extracting geometrical profiles of nanopipettes (\txblBtoBl{SI\,7} and \txblRtoBl{\txblBtoBl{Figure S7}}), \txblRtoBl{and finite element method simulations and experiments for the estimation of relaxation rate of ion concentration gradient in the tip (\txblBtoBl{SI\,8} and \txblBtoBl{Figure S8}, \txblBtoBl{Figure S9}, and \txblBtoBl{Figure S10})} (\txblBtoBl{PDF}).\\
    Video files are also provided for procedure of cutting nanopipette (\txblBtoBl{Movie 1}).\\
    Procedure for position adjustment of tip segment on TEM grid (\txblBtoBl{Movie 2}).\\
    Dynamics of nanobubbles confined in sub-10-nm nanopipette (\txblBtoBl{Movie 3}).\\

    % The Supporting Information is available free of charge on the ACS Publications website.
    % Pulling parameters, optimized pulling condition, photograph of nanopipettes on the TEM grid, for sub-10-nm nanopipettes, TEM images of sub-10-nm nanopipettes with the membrane, procedures for abstracting the shape profile of  nanopipettes, and Dynamics of nanobubbles in sub-10-nm nanopipettes (\textcolor{blue}{PDF})\\
    % Procedure for cut of nanopipettes (\textcolor{blue}{Movie 1})\\
    % Procedure for alignment and fixation of nanopipette tip on grids (\textcolor{blue}{Movie 2})\\
    % Dynamics of nanobubbles in sub-10-nm nanopipettes (\textcolor{blue}{Movie 3})
    %%%%%%%%%%%%%%%%%%%%%%%%%%%%%%%%%%%%%%%%%%%%%%%%%%%%%%%%%%%%%%%%%%%%%
\subsubsection*{AUTHOR INFORMATION}
    \subsubsection*{Corresponding Authors}\mbox{}\\
    % E-mail: wshinji@se.kanazawa-u.ac.jp (S.W).\\
    % E-mail: tando@staff.kanazawa-u.ac.jp (T.A).\\ 
    \noindent
    Shinji Watanabe,~phone +81(0)76 2344054\\
    \noindent
    Toshio Ando, ~phone +81(0)76 2645663\\ 
    
    \subsubsection*{ORCID}\mbox{}\\
    Kazuki Shigyou:0000-0003-4195-3628\\
    Linhao Sun:0000-0002-4325-7303\\
    Yousuke Kikuchi:0000-0002-2278-821X\\
    Hirotoshi Furusho:0000-0002-6286-5304\\
    Keisuke Miyazawa:0000-0002-5012-8040\\
    Takashi Fukuma:0000-0001-8971-6002\\
    Azuma Taoka:0000-0003-2492-7858\\
    %Shinji Watanabe:0000-0002-4831-0185\\
    Toshio Ando:0000-0001-8819-154X\\
    Shinji Watanabe:0000-0002-4831-0185\\

%\begin{acknowledgement}
\begin{acknowledgements}

    This work was supported by a grant of JST SENTAN (JPMJSN16B4 to S.W.), Grant for Young Scientists from HOKURIKU Bank (to S.W.), JSPS Grant-in-Aid for Young Scientists (B) (JP26790048 to S.W.), JSPS Grant-in-Aid for Young Scientists (A) (JP17H04818 to S.W.), JSPS Grant-in-Aid for Scientific Research on Innovative Areas (JP16H00799 to S.W.) and JSPS Grant-in-Aid for Challenging Exploratory Research (JP18K19018 to S.W.), and JSPS Grant-in-Aid for Scientific Research (S) (JP17H06121 and JP24227005 to T.A.). This work was also supported by a Kanazawa University CHOZEN project and World Premier International Research Center Initiative (WPI), MEXT, Japan.

%\end{acknowledgement}
\end{acknowledgements}

%\bibliography{achemso-demo.bib}
%! ACSにアップするときは、同じディレクトリにbibファイルを置くべし。
\bibliography{reference_20141021.bib}
%\bibliography{../../../Biblio/reference_20141021.bib}
%\bibliography{KS(TEM)_200311.bbl}

% \clearpage
% \section*{Graphical TOC Entry}
% \vspace{2cm}
% %\begin{center}
% \begin{figure}[H]
%     \includegraphics{TOC.pdf}
%     \label{TOC}
%     \end{figure}
% %\end{center}

\end{document}